\documentclass[12pt]{iopart}
\usepackage{iopams,amstext,graphicx,dsfont}
\begin{document}

\title[Lightcone renormalization and quantum quenches]{Lightcone
  renormalization and quantum quenches in one-dimensional Hubbard
  models}

\author{Tilman Enss}
\address{Physik Department, Technische Universit\"at M\"unchen,
  D-85747 Garching, Germany}
\author{Jesko Sirker} 
\address{Department of Physics and Research Center OPTIMAS,
  Technical University Kaiserslautern, D-67663 Kaiserslautern,
  Germany}

\date{\today}

\begin{abstract}
  The Lieb-Robinson bound implies that the unitary time evolution of
  an operator can be restricted to an effective light cone for any
  Hamiltonian with short-range interactions. Here we present a very
  efficient renormalization group algorithm based on this light cone
  structure to study the time evolution of prepared initial states in
  the thermodynamic limit in one-dimensional quantum systems.  The
  algorithm does not require translational invariance and allows for
  an easy implementation of local conservation laws. We use the
  algorithm to investigate the relaxation dynamics of double
  occupancies in fermionic Hubbard models as well as a possible
  thermalization. For the integrable Hubbard model we find a pure
  power-law decay of the number of doubly occupied sites towards the
  value in the long-time limit while the decay becomes exponential
  when adding a nearest neighbor interaction. In accordance with the
  eigenstate thermalization hypothesis, the long-time limit is
  reasonably well described by a thermal average. We point out though
  that such a description naturally requires the use of negative
  temperatures. Finally, we study a doublon impurity in a N\'eel
  background and find that the excess charge and spin spread at
  different velocities, providing an example of spin-charge separation
  in a highly excited state.
\end{abstract}


\maketitle
\tableofcontents

\section{Introduction}
\label{Sec_Intro}
Using ultracold atomic gases as quantum simulators, it has become
possible to prepare states in almost perfectly isolated many-body
systems and to monitor their time evolution \cite{KinoshitaWenger,
  HofferberthLesanovsky, GerickeWuertz, WuertzLangen, StrohmaierGreif,
  SensarmaPekker}.  At the same time, enormous progress in numerical
renormalization group methods has given us access to the dynamics of
quantum models in one dimension (1D) \cite{DaleyKollath, FeiguinWhite,
  SirkerKluemperDTMRG, SirkerPereira, SirkerPereira2, VidalTEBD1,
  VidaliTEBD}.  These algorithms are all based on approximating a
quantum state as a matrix product in an optimally chosen truncated
Hilbert space, an idea dating back to the density matrix
renormalization group (DMRG) by White \cite{WhiteDMRG}.  This makes it
now possible to study, both experimentally and numerically,
fundamental questions about the relaxation dynamics and the role of
conservation laws \cite{KinoshitaWenger, RigolDunjko}.  Furthermore,
the applicability of the eigenstate thermalization hypothesis
(ETH)---according to which each generic state of a closed quantum
system already contains a thermal state which is revealed during
unitary time evolution by dephasing \cite{Deutsch, Srednicki,
  CassidyClark}---can be investigated as well.

\begin{figure}[!ht]
  \begin{center}
    \includegraphics*[width=.65\linewidth]{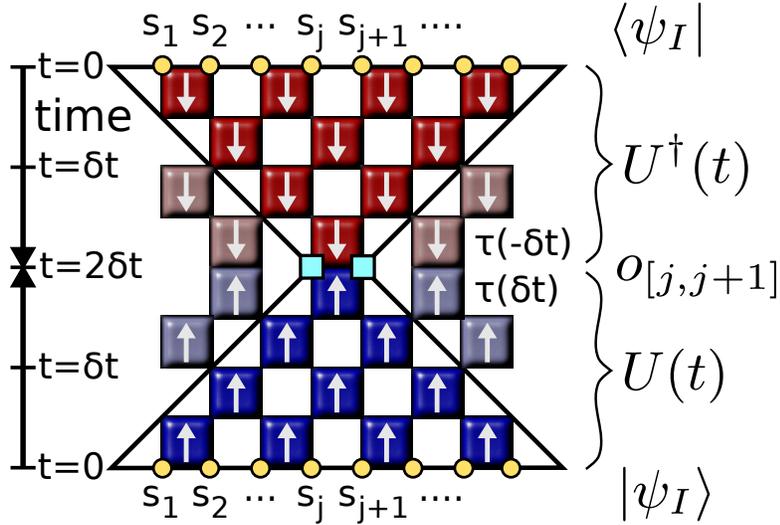}
  \end{center}
  \caption{\label{FigTrotter} LCRG algorithm.  The Trotter-Suzuki
    decomposition of time evolution reveals the light cone structure.}
\end{figure}
In Sec.~\ref{Sec_LCRG} we present a new algorithm to study the unitary
time evolution of an initial state in a 1D quantum system. We
concentrate on the case of a product initial state particularly
relevant for experiment but note that the algorithm has been
implemented also for thermal initial states.  The main idea is to make use
of the Lieb-Robinson bound \cite{LiebRobinson} to efficiently simulate
the system and to obtain results directly in the thermodynamic
limit. Let us briefly recapitulate one of the main results of
Refs.~\cite{LiebRobinson, BravyiHastings} which is the basis for our
algorithm. We are interested in the time evolution of quantum systems
starting from some initial state $|\Psi_I\rangle$ where all connected
correlation functions decay exponentially with a finite correlation
length $\xi$.  The time evolution of a local operator $o_{[j,j+n]}$
acting on sites $j,j+1,\ldots, j+n$ can then be approximated by an
operator acting only in the effective light cone of the region
$[j,j+n]$ (see Fig.~\ref{FigTrotter}) while being the identity
operator outside of the light cone. More precisely, if
$o^l_{[j,j+n]}(t)$ is the time evolved operator active only on sites
which are at most distance $l$ apart from the region $[j,j+n]$ then
\begin{equation}
  \label{LiebRobinson}
  ||o_{[j,j+n]}(t) -o^l_{[j,j+n]}(t)|| 
  \leq \text{const}\times \exp\left(-\frac{l-v_{LR}|t|}{\xi}\right)
\end{equation}
where $v_{LR}$ is the Lieb-Robinson velocity which is typically, in
natural units, of the order of the interaction parameters of the model
under consideration \cite{BravyiHastings}. If $v_{LR}|t|\ll l$ then
the error of approximating $o_{[j,j+n]}(t)$ by $o^l_{[j,j+n]}(t)$ is
exponentially small.  We show in Sec.~\ref{Sec_LCRG} that a
Trotter-Suzuki decomposition of unitary time evolution immediately
leads to a light cone and that this light cone can be represented in a
truncated Hilbert space using density matrix renormalization group
(DMRG) techniques \cite{WhiteDMRG, BursillGehring, WangXiang,
  KemperSchadschneider, EnssSchollwoeck, KemperGendiar, EnssHenkel,
  SirkerKluemperDTMRG, Hastings, Banuls}.  In contrast to ground state
and transfer matrix DMRG algorithms an explicit calculation of
eigenvectors of the system, which is the computationally most costly
step, is not necessary. This makes the new light cone renormalization
group (LCRG) algorithm extremely fast and efficient.  Furthermore, the
implementation of local conservation laws---important for an effective
numerical study---becomes particularly simple. The ``speed of light''
set by the Trotter-Suzuki decomposition is typically chosen to be much
larger than the Lieb-Robinson speed $v_{LR}$ at which information
spreads so that the algorithm directly yields results for the
thermodynamic limit.  Contrary to the infinite size time evolving
block decimation (iTEBD) \cite{VidaliTEBD}, however, it does not rely
on translational invariance. In our paper we will demonstrate these
advantages of the LCRG algorithm by studying several examples. At the
same time we note that our approach does not solve the most
fundamental problem of using matrix product states to investigate the
time evolution of one dimensional quantum systems: the linear growth
of entanglement entropy with time, which restricts the applicability
of such methods to the intermediate time dynamics. It can be shown
under very general conditions that this is a fundamental property of
unitary time evolution \cite{LiebRobinson, BravyiHastings} which
cannot easily be overcome.

In Sec.~\ref{Sec_Hubb} we will apply the LCRG algorithm to study the
relaxation of a doublon lattice in 1D fermionic Hubbard models.  While
the problem of a single doublon-holon pair has already been studied in
1D \cite{Alhassanieh}, our study is mainly motivated by the
experimental and theoretical investigation of the decay of a
macroscopic number of doublons in an ultracold fermionic gas on a
three-dimensional optical lattice \cite{StrohmaierGreif, RoschRasch}.
First, we will present a test of the algorithm by studying the free
fermion case where the time evolution can be calculated analytically.
Next, we will investigate the differences in the relaxation dynamics
between the interacting integrable and non-integrable cases as well as
a possible thermalization in the long-time limit. In
Sec.~\ref{Sec_Non-trans} we will then demonstrate one of the major
advantages of the LCRG algorithm: Even for systems without
translational invariance, results in the thermodynamic limit can be
obtained. In Sec.~\ref{Conc} we give a brief summary and an outlook on
possible future applications of the algorithm. The supplementary
material contains the executable code of the LCRG for the anisotropic
Heisenberg model, which we have chosen as a simple example, as well as
videos of the time evolution for the problem studied in
Sec.~\ref{Sec_Non-trans}.

\section{The light cone renormalization group algorithm}
\label{Sec_LCRG}
We present the LCRG algorithm to compute the time evolution 
\begin{equation}
  \label{Dstate}
  \langle o_{[j,j+n]}\rangle^{I}(t)
  \equiv \langle\Psi_I| \e^{\rmi H t}\, o_{[j,j+n]}\, \e^{-\rmi H t} |\Psi_I\rangle 
\end{equation}
of a local operator $o_{[j,j+n]}$ acting on sites $j,j+1,\ldots, j+n$.
For the initial state $|\Psi_I\rangle = |s_1\, s_2\, \ldots \rangle$
we consider a product state with $s_j$ denoting states in the local
basis of dimension $M$.  We note that with the help of ancilla sites
also thermal states can be expressed using a product initial state
followed by an imaginary time evolution, which is also performed using
the light-cone algorithm.  In this way we have implemented the real
time evolution starting, e.g., from a highly entangled quantum state
such as the ground state.
We consider a Hamiltonian $H=\sum_j h_{j,j+1}$ with nearest neighbor
interaction; a Trotter-Suzuki decomposition of the unitary time
evolution operator then leads to the 2D lattice shown graphically in
Fig.~\ref{FigTrotter}.  It consists of local updates of two
neighboring sites forward in time $\tau_{j,j+1}(\delta
t)=\exp(-ih_{j,j+1}\delta t)$ (``$\uparrow$'' plaquettes) and backward
in time $\tau_{j,j+1}(-\delta t)\equiv\tau^\dagger_{j,j+1}(\delta t)$
(``$\downarrow$'' plaquettes), where $\delta t$ is the Trotter-Suzuki
time step.  Unless there is an operator insertion, facing plaquettes
trivialize and become the identity operator, $\tau_{j,j+1}(-\delta
t)\, \tau_{j,j+1}(\delta t) = \mathds{1}$ (shaded plaquettes).  This
yields the light cone structure emanating from the local observable
$o_{[j,j+n]}$ at time $t$.  As long as the ``speed of light'' of the
Trotter-Suzuki decomposition is larger than the Lieb-Robinson velocity
$v_{LR}$ the expectation value \eref{Dstate} is effectively evaluated
in the thermodynamic limit. Neither translational invariance of the
initial state nor of the Hamiltonian are required for this
construction.

\begin{figure}[!ht]
  \begin{center}
    \includegraphics*[width=.4\linewidth]{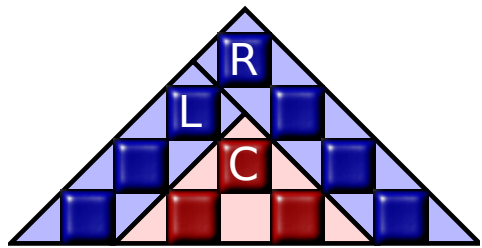}
  \end{center}
  \caption{\label{FigAlg} LCRG algorithm.  The light cone $C$ grows
    with each time step by adding first a diagonal left transfer
    matrix $L$ and then a diagonal right transfer matrix $R$.}
\end{figure}

The LCRG algorithm is based on corner transfer matrices \cite{Baxter,
  NishinoOkunishi1} to compute the growth of the light cone with each
successive time step $\delta t$ (Fig.~\ref{FigAlg}): the light cone
$C_t$ at time $t$ is multiplied from the left with the diagonal left
transfer matrix $L$ and then from the right with the diagonal right
transfer matrix $R$ to construct the new light cone for the next time
step, $C_{t+\delta t}$.  Of course a direct implementation of this
procedure would quickly break down because the Hilbert space of light
cone states grows exponentially with time.  Therefore, we use ideas
from DMRG studies of dynamics in stochastic systems
\cite{KemperGendiar, EnssHenkel} to represent both the light cone $C$
and the transfer matrices $L$, $R$ in a reduced Hilbert space of
manageable dimension.  A fully working implementation of this
algorithm specialized to homogeneous systems in included in the
supplementary material.

\begin{figure}[!ht]
  \begin{center}
    \includegraphics*[width=.65\linewidth]{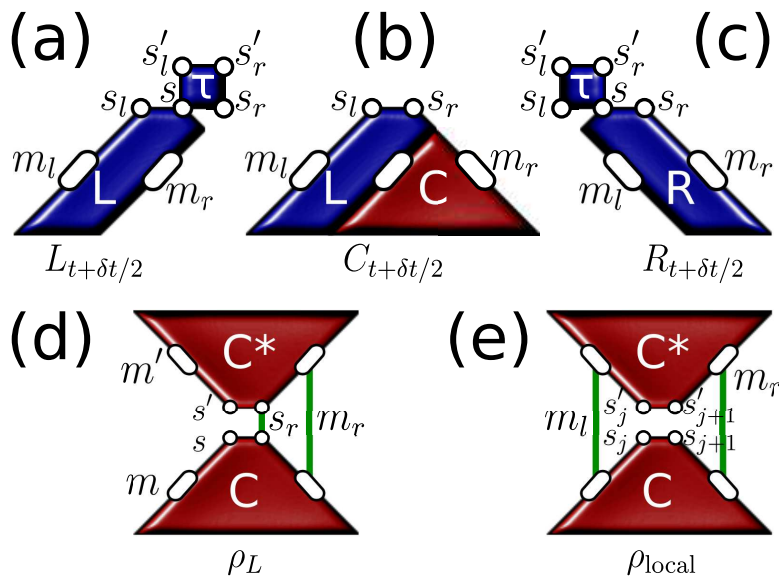}
  \end{center}
  \caption{\label{FigSteps} LCRG algorithm.  (a)-(c) The light cone
    $C$ grows to the left by contraction with a left transfer matrix
    $L$; the left and right transfer matrices $L$ and $R$ are
    augmented by a local plaquette $\tau(\delta t)$.  (d) The reduced
    density matrix $\rho_L$ is constructed from the forward and
    backward light cones by tracing only over the right site and block
    indices.  (e) The local density matrix is obtained by tracing the
    forward and backward light cones over both left and right block
    indices.}
\end{figure}

In practice the time evolution proceeds in two half time steps (see
Fig.~\ref{FigAlg}). In the first step, the light cone $C_t[m_lm_r]$
has left and right block indices (Fig.~\ref{FigSteps}b) representing
states in the (reduced) Hilbert space of dimension $\chi$, while the
left transfer matrix $L_t[m_ls_lm_rs_r]$ has again two block indices
but also left and right site indices $s_l$, $s_r$ of dimension $M$
(Fig.~\ref{FigSteps}a).  $L_t$ and $C_t$ are contracted over their
common block index to yield the new light cone $C_{t+\delta t/2}$ half
a time step ahead (Fig.~\ref{FigSteps}b):
\begin{equation}
  \label{eq:step1}
  C_{t+\delta t/2}[m_ls_ls_rm_r] = \sum_m L_t[m_ls_lms_r]\, C_t[mm_r] \,.
\end{equation}
The left transfer matrix $L_t$ is enlarged by adding a plaquette at
its upper right site index $s_r$ (Fig.~\ref{FigSteps}a):
\begin{equation}
  \label{eq:step2}
  L_{t+\delta t/2}[m_ls_ls_l'm_rs_rs_r'] = \sum_s L_t[m_ls_lm_rs]\,
  \tau[s_l's_r'ss_r] \,.
\end{equation}
Similarly, a local plaquette is attached to the upper left corner of
the right transfer matrix $R_t$ to construct $R_{t+\delta t/2}$
(Fig.~\ref{FigSteps}c).  The initial conditions at $t=0$ are as
follows: the block indices represent a single site with dimension $m_l
= m_r = M$, the intial light cone is $C_{t=0}[m_l m_r] = \Psi_I[m_l
m_r]$ for the product initial state $|\Psi_I\rangle$ on two
neighboring sites, and the transfer matrices have the initial forms
$L_{t=0}[m_l s_l m_r s_r] = \sum_s \Psi_I[m_l s]\, \tau[s_l s_r s
m_r]$ and $R_{t=0}[s_l m_l s_r m_r] = \sum_s \tau[s_l s_r m_l s]\,
\Psi_I[s m_r]$.

In order to bring $C$, $L$ and $R$ back into their original form the
old block index $m$ (dimension $\chi$) is combined with the adjacent
site index $s$ (dimension $M$) into a new block index $m'=(ms)$ of
dimension $\chi' = M\chi$.  The challenge is to limit the exponential
growth of $\chi$ with every time step.  This is done by a
renormalization step where a reduced density matrix is used to select
the $\chi$ most important basis states within the $\chi'$-dimensional
Hilbert space.  The reduced density matrix $\rho_L$ for the left block
index is formed by combining the forward and backward light cones and
tracing over the right site and block indices (Fig.~\ref{FigSteps}d)
\begin{equation}
  \label{eq:step3}
  \rho_L[(m's')(ms)] = \sum_{s_rm_r} C_{t+\delta t/2}^*[(m's')s_rm_r]\,
  C_{t+\delta t/2}[(ms)s_rm_r] 
\end{equation}
where we have used the fact that in unitary time evolution the
backward light cone is the adjoint of the forward light cone.  The
reduced density matrix $\rho_L$ of dimension $\chi'$ is by
construction hermitean and has unit trace.  $\rho_L$ is diagonalized,
and the $\chi$ states with the largest eigenvalues form the basis of
the reduced Hilbert space. Optionally, one can retain all states such
that the cumulative weight of the discarded states remains below a
given threshold.  We use a combination of both to obtain a reliable
error control. Finally, the
left block index of the light cone $C$ and both block indices of the
left transfer matrix $L$ are projected onto this reduced basis, $(m_ls_l)
\mapsto m_l$.  Analogously, the reduced density matrix $\rho_R$ is
formed by tracing over the left block indices to find a reduced basis
for the right block indices, and subsequently the right block index of
$C$ and both block indices of $R$ are projected onto the reduced
Hilbert space.  This completes the first half time step.

The second half of the algorithm works similarly by joining a right
transfer matrix $R$ to the right of the light cone
(Fig.~\ref{FigAlg}),
\begin{equation}
  \label{eq:step4}
  C_{t+\delta t}[m_ls_ls_rm_r] = \sum_m C_{t+\delta t/2}[m_lm]\,
  R_{t+\delta t/2}[s_lms_rm_r] \,.
\end{equation}
At this stage the local density matrix
\begin{equation}
  \label{eq:rholocal}
  \rho_\text{local}(t+\delta t)[s_j's_{j+1}'s_js_{j+1}] = \sum_{m_lm_r}
  C_{t+\delta t}^*[m_ls_j's_{j+1}'m_r]\, C_{t+\delta t}[m_ls_js_{j+1}m_r]
\end{equation}
is formed by contracting the forward and backward light cones over the
left and right block indices, leaving open the site indices in the
middle (Fig.~\ref{FigSteps}e).  The expectation value of a local
operator $o_{[j,j+n]}$ is then obtained as
\begin{equation}
  \label{eq:explocal}
  \langle o_{[j,j+1]} \rangle^I(t+\delta t) = \Tr_{[j,j+1]} \Bigl(
    \rho_\text{local}(t+\delta t)\, o_{[j,j+1]} \Bigr) .
\end{equation}
By multiplying further transfer matrices onto the left or right one can
form also the local density matrix $\rho_{[j,j+n]}$ spanning more than
two neighboring sites.  For example, the density profile to the left
of an impurity site is obtained by starting with $\rho_L$ and
repeatedly multiplying $L$ from the left onto the lower light cone and $L^*$ onto
the upper light cone, until the desired distance from the impurity is
reached.  The remaining second half time step proceeds in
complete analogy with the first part, growing $L$ and $R$ by one
plaquette and renormalizing in turn the left and right block indices.

Note that only summations and multiplications are required to build
the light cone.  This saves the most time-consuming step in standard
transfer matrix DMRG algorithms, where one has to find the largest
eigenvector of the transfer matrix.  Only the density matrix
$\rho_{L,R}$ has to be diagonalized, which dominates the computation
time $\mathcal O(M^3 \chi^3)$. Our algorithm therefore combines the
speed of iTEBD \cite{VidaliTEBD} with the flexibility of TEBD
\cite{VidalTEBD1} to treat non-translationally invariant systems.  Due
to the local structure of the updates, conservation laws are easily
implemented in our algorithm (see below).  We note that instead of the
first order Trotter-Suzuki decomposition shown in
Fig.~\ref{FigTrotter} also higher order decompositions can be easily
implemented.

\section{Doublon decay in Hubbard models}
\label{Sec_Hubb}
We use the LCRG algorithm with a second order Trotter-Suzuki
decomposition to study dynamics in the 1D fermionic Hubbard model
\begin{eqnarray}
  \label{Hubbard}
  H_{U,V} = & -J\sum_{j,\sigma=\uparrow,\downarrow}
  (c^\dagger_{j,\sigma} c_{j+1,\sigma} + h.c.)
  + U \sum_j
  \left(n_{j\uparrow}-\frac{1}{2}\right)\left(n_{j\downarrow}-\frac{1}{2}\right)
  \nonumber \\
  & + V\sum_j (n_j-1)(n_{j+1}-1)
\end{eqnarray}
where $J$ is the hopping amplitude,
$n_{j,\sigma}=c^\dagger_{j,\sigma}c_{j,\sigma}$ and $n_j =
n_{j\uparrow}+n_{j\downarrow}$ the occupation numbers, $U$ the onsite,
and $V$ the nearest-neighbor potential.
As initial states we will consider 
the state $|\Psi_D\rangle$, where doubly occupied and empty sites
alternate, and the N\'eel state, $|\Psi_N\rangle$. These state are
given explicitly by
\begin{eqnarray}
\label{states}
|\Psi_D\rangle =\prod_j
c^\dagger_{2j\uparrow}c^\dagger_{2j\downarrow}|0\rangle\,, \qquad && \qquad
|\Psi_N\rangle = \prod_j
c^\dagger_{2j+1\uparrow}c^\dagger_{2j\downarrow}|0\rangle 
\end{eqnarray}
where $|0\rangle$ denotes the vacuum.  For these states, we want to
investigate the time dependence of 
double occupancies, $d_j=n_{j\uparrow}n_{j\downarrow}$, the staggered
magnetization,
$m_j=(-1)^jS^z_j=(-1)^j(n_{j\uparrow}-n_{j\downarrow})/2$, and of the
operator $w_j=(-1)^j n_j$, measuring the charge imbalance between even
and odd sites.  Before discussing the numerical results, we first want to
establish a number of relations between these three operators
in the case where the nearest neighbor repulsion
vanishes, $V=0$. The model with $V\neq 0$ will be studied in
Sec.~\ref{Sec_ExtHubb}.
%% which follow from the symmetries of the Hubbard model.

\subsection{Duality relations for the integrable model}
For $V=0$ the model (\ref{Hubbard}) becomes the integrable Hubbard
model.  Apart from the special symmetries responsible for the
integrability of the model by Bethe ansatz, there are other symmetries
in this case %% The following symmetries of the Hubbard model
which allow us to establish various relations between the states and
operators: \\
(a) There is a unitary duality transformation
\begin{equation}
  \label{duality}
  \mathcal{U}=\prod_j \Bigl(c_{j\uparrow} + (-1)^jc^\dagger_{j\uparrow}\Bigr)
\end{equation}
relating the repulsive ($U>0$) and the attractive ($U<0$) Hubbard
models. This transformation leads to $\mathcal{U}^\dagger
c_{j\uparrow}\mathcal{U} = (-1)^jc^\dagger_{j\uparrow}$, $
\mathcal{U}^\dagger c_{j\downarrow}\mathcal{U} = c_{j\downarrow} $ so
that the kinetic energy part in Eq.~(\ref{Hubbard}) stays invariant
while $U\to -U$. For the operators we find 
\begin{eqnarray}
  d_j&\to &n_{j\downarrow}-d_j,\\
  m_j&\to &(-1)^j(1-n_j)/2, \\
  w_j&\to &(-1)^j(1+n_{j\downarrow}-n_{j\uparrow}).
\end{eqnarray}
For the initial state it follows that
$\mathcal{U}^\dagger|\Psi_D\rangle = |\Psi_N\rangle$,
$\mathcal{U}^\dagger|\Psi_N\rangle = |\Psi_D\rangle$, assuming an even
number of lattice sites $L$. For the expectation values (see also Eq.~(\ref{Dstate}))
\begin{equation}
  o_U^I(t)\equiv\sum_{j=1}^L\langle o_j\rangle^I_U(t)/L\, ,
\end{equation}
the duality transformation implies
\begin{eqnarray}
  \label{eq:dUD}
  d_U^D(t)&=&1/2-d_{-U}^N(t),\\
  \label{eq:mUN}
  m_U^N(t)&=&-w_{-U}^D(t)/2,\\
  \label{eq:mUD}
  m_U^D(t)&=&-w_{-U}^N(t)/2\equiv 0.
\end{eqnarray}
The expectation values in the last equation \eref{eq:mUD} have to
vanish identically because of the particle-hole (spin inversion)
symmetry of the initial states, respectively. The second identity
\eref{eq:mUN}, furthermore, shows that the decay of the staggered
magnetization can be studied in a realization of a fermionic Hubbard
in cold atomic gases without the need to address the spin degree of
freedom directly in a measurement.\\
(b) On a bipartite lattice, $A\otimes B$, we can furthermore apply the
transformation $c_{j\sigma}\to \pm c_{j\sigma}$ for $j\in A$ ($j\in
B$), respectively. This leads to $J\to -J$, $U\to U$ and therefore
$H_{-U}\to -H_U$.
This results in $d_U^D(t)=1/2-d_{U}^N(-t)$ and similarly for the other
identities.\\
(c) Finally, we can use the time reversal invariance of the
expectation values. Using all three symmetries we find
\begin{eqnarray}
  d_U^D(t)&=&d_{-U}^D(t)=1/2-d_{U}^N(t), \\
  m_U^N(t)&=&m_{-U}^N(t)=-w_U^D(t)/2.
\end{eqnarray}
The relaxation dynamics we will consider here is therefore independent
of the sign of $U$ and the same information is obtained by starting
either from $|\Psi_D\rangle$ or $|\Psi_N\rangle$.

\subsection{Testing the LCRG algorithm: The free fermion case}
To test the LCRG algorithm, we first study the free spinful fermion
(SFF) case $U=0$ where the dynamics can be calculated exactly. We find
$d_{U=0}^D(t)=(1+J^2_0(4Jt))/4$ and $m_{U=0}^N(t) = J_0(4Jt)/2$ with
$J_0$ the Bessel function of the first kind.
\begin{figure}[t!]
\begin{center}
\includegraphics*[width=0.7\columnwidth]{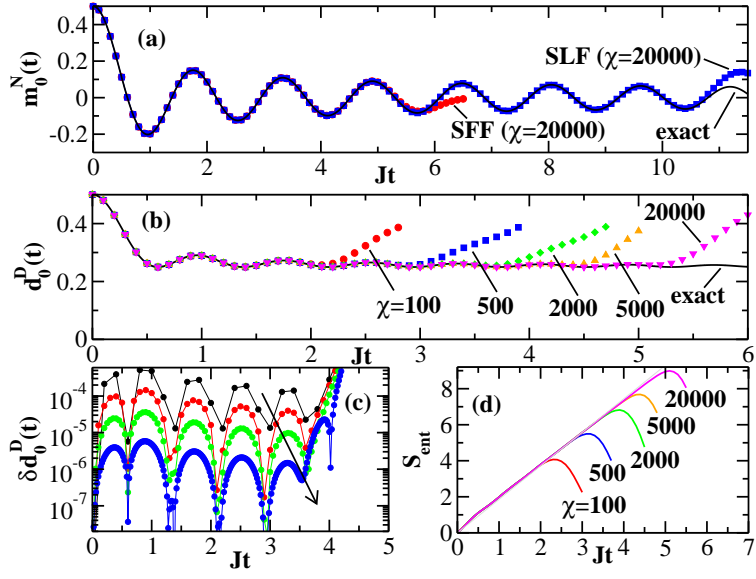}
\end{center}
\caption{(a) $m_0^N(t)$ for free SFF (circles) and for free SLF
  (squares)---note that the time scale in \cite{BarmettlerPunk} is
  stretched by a factor $2$. (b) $d_0^D(t)$ for free SFF with $\chi$
  as indicated.  In both cases $\delta t=0.1$ and lines denote the
  exact results. (c) Absolute error $\delta d_0^D(t)$ for free SFF
  with $\chi=5000$ and $\delta t =0.2,0.1,0.05,0.02$ (in arrow
  direction).  (d) $S_{\rm ent}(t)$ for free SFF with $\delta t =0.1$
  and $\chi$ as indicated.}
\label{Fig1}
\end{figure}
In the free SFF case, the dynamics of electrons with spin up and spin
down is completely decoupled. Therefore, we can also use free spinless
fermions (SLF) to calculate $m_0^N(t)$ with a spinless particle
representing either the presence of a spin up or a spin down. Then we
need to keep only $\sqrt{\chi}$ states to simulate the dynamics with
the same accuracy.  In Fig.~\ref{Fig1}(a) the LCRG results for
$m_0^N(t)$ for free SFF and SLF are compared to the exact result. For
free SLF with $\chi=20000$ block states we are able resolve $6.5$
oscillations compared to the $5$ oscillations which have been resolved
in \cite{BarmettlerPunk} by iTEBD.  We emphasize that for the Hubbard
model ($M=4$) with the conservation laws for spin and charge
implemented and $\chi=2000$ states kept, each time step takes only
$\sim 30$ seconds on a standard PC without parallelization. This is
$260\times$ faster and uses $12\times$ less memory than without
conservation laws, because the largest diagonal block of the reduced
density matrix $\rho_{L,R}$ has only $200$ states.  For SLF ($M=2$)
the speedup is still $40\times$ with $5\times$ less memory and a
largest block of $450$ states.  In Fig.~\ref{Fig1}(b) the results for
$d_{0}^D(t)$ are shown where $\chi$ is varied. The error of the
simulation up to $t_{\rm max}$ where the simulation starts to deviate
from the exact result is completely dominated by the error of the
Trotter-Suzuki decomposition (see Fig.~\ref{Fig1}(c)) and is of order
$(\delta t)^2$ for the second order decomposition used
here. Importantly, $t_{\rm max}$ is determined only by $\chi$ and
results with in principle arbitrary accuracy can be obtained for $t\in
[0,t_{\rm max}]$ by reducing the time step $\delta t$ or using a
higher order Trotter-Suzuki decomposition, since the number of RG
steps is not restricted.

The algorithm breaks down when the spectrum of the reduced density
matrix $\rho_s=\rho_{L,R}$ becomes dense. A suitable measure is the
entanglement entropy
\begin{equation}
 S_{\rm ent}(t) =-\mbox{Tr}\, \rho_s\ln\rho_s\leq
\ln(M\chi)
\end{equation}
with $M\chi=\mbox{dim}\,\rho_s$. The entanglement entropy is shown in
Fig.~\ref{Fig1}(d) and increases linearly with time. We want to remind
the reader once more that the linear increase of the entanglement
entropy seems to be a fundamental property of unitary time evolution
\cite{BravyiHastings} which cannot easily be overcome and limits the
simulation time.

The LCRG algorithm actually does provide an intuitive picture for this
behavior: Facing plaquettes outside of the light cone (shown shaded in
Fig.~\ref{FigTrotter}) trivialize, thereby connecting a local degree
of freedom at the edge of the lower light cone with one on the upper
light cone by a Kronecker delta. The number of these Kronecker delta
bonds between the lower and upper light cone increases linearly with
time and determines the entanglement entropy between the light cones.
This is very similar to the entanglement entropy of a spin-$1/2$
Heisenberg chain: In this case the ground state can be represented in
a resonating valence bond (RVB) basis. If the chain is now split into
two semi-infinite segments then the entanglement entropy has been
shown to be proportional to the number of RVB bonds connecting the
segments \cite{AletCapponi}.  A breakdown of the simulation is
observable as a deviation from the linear growth of $S_\text{ent}$ and
occurs when the entanglement entropy is close to the bound, $S_{\rm
  ent}(t)\sim\ln(M\chi)$, i.e., when all eigenvalues of $\rho_s$ have
comparable magnitude thus making further RG steps impossible.

\subsection{Results for the Hubbard model}
\begin{figure}[t!]
\begin{center}
\includegraphics*[width=0.7\columnwidth]{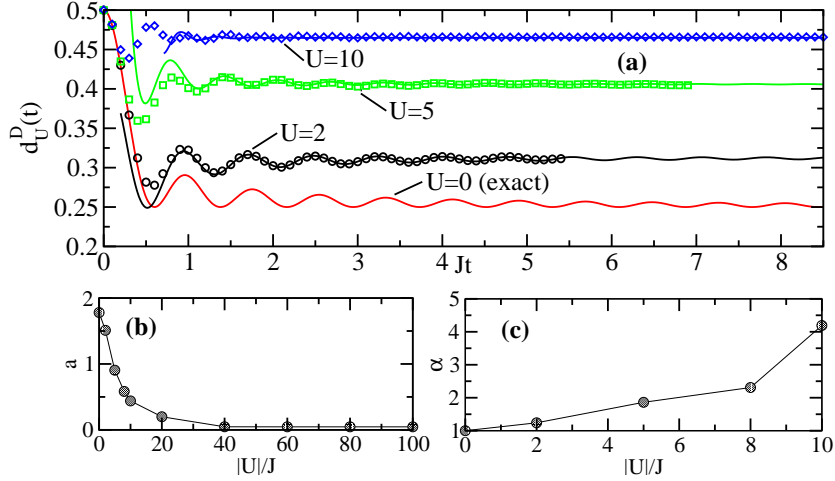}
\end{center}
\caption{Evolution of double occupancy in Hubbard model. 
  (a) $d^D_U(t)$ with $\chi =20000$, $\delta
  t=0.1$ (symbols), and fits (lines), see text.  (b) Slope of the
  entanglement entropy, $S_{\rm ent} \sim aJt$. (c) Fitted exponent
  $\alpha$ of the power-law decay, see Eq.~(\ref{fit_fct}).}
  \label{Fig2}
\end{figure}

Next, we study $d^D_U(t)$ in the interacting Hubbard model, a
situation which can be realized in ultracold gases
\cite{StrohmaierGreif}.  For times $Jt\ll \min\{J/|U|,1\}$ the
relaxation is independent of the interaction strength and follows the
short-time expansion of the free fermion result $d_0^D(t)\sim 1/2
-2(Jt)^2$, see Fig.~\ref{Fig2}(a).  Thus, in order to see the effect
of interactions, systems at times $Jt\gg 1/|U|$ have to be studied.
In units of the hopping amplitude $J$ we can simulate longer times the
larger $U$ is. This is a consequence of the slower increase of $S_{\rm
  ent}(t) \sim aJt$ as shown in Fig.~\ref{Fig2}(b). For large $U$ we
find that the slope of the entanglement entropy is given by $a\sim
J/|U|$, i.e., the simulation time is proportional to $t_{\rm max}\sim
|U|/J^2$ and therefore set by the inverse of the magnetic
superexchange interaction $\sim J^2/|U|$. It is clear that the slope
of the entanglement growth becomes smaller the closer the initial
state is to an eigenstate of the Hamiltonian: the eigenstate stays
invariant under time evolution and no additional entanglement entropy
is generated. Comparatively long times can therefore be simulated, in
particular, if the time evolution of the ground state with a weak
perturbation is studied as, for example, in Ref.~\cite{Alhassanieh}.

At times $Jt\gg 1$ the relaxation in the free SFF case is given by
$d^D_0(t)=(1+J^2_0(4Jt))/4\sim [1+(4\pi
t)^{-1}(1+\cos(8Jt-\pi/2))]/4$.  This motivates us to fit the
time dependence at finite $U$ by the function
\begin{equation}
\label{fit_fct}
d^D_U(t) = d_U^D(\infty) +e^{-\gamma t}[\mathcal{A}+\mathcal{B}\cos(\Omega t-\phi)]/t^\alpha 
\end{equation}
in the regime $1.5 < Jt \leq Jt_{\rm max}$. Such fits are shown as
solid lines in Fig.~\ref{Fig2}(a). In all cases $\gamma < 10^{-3}$,
i.e., we do not find evidence for a finite relaxation rate $\gamma$
\footnote{We note that the fits for large $|U|$ are more ambiguous
  because $d_U^D(\infty)$ is reached more quickly.}. A relaxation
following a power law has also been observed at intermediate times in
a 1D Bose-Hubbard model starting from an initial state with one boson
on every second site \cite{CramerFlesch}. On the other hand, the
relaxation for the $XXZ$ model when starting from a N\'eel state has
been interpreted in terms of an exponential decay
\cite{BarmettlerPunk}.  Our fits point to a pure power-law decay with
an exponent $\alpha$ which increases with increasing $|U|$, see
Fig.~\ref{Fig2}(c). The asymptotic value $d^D_U(\infty)$ also
increases and reaches $1/2$ in the limit $|U|\to\infty$, see the $V=0$
data in Fig.~\ref{Fig4b}(b) below.  We emphasize that
$d^D_U(t)=d^D_{-U}(t)$, i.e., repulsive interactions lead to a binding
of doublons the same way as attractive interactions do \cite{Winkler}.
For doublons moving on a 3D lattice it has been argued that for
repulsive interactions, $U>0$, much larger than the bandwidth,
many-body scattering processes are needed to dissipate the doublon
energy, which leads to an exponentially small relaxation rate
$\gamma\sim\exp(-U/J)$ \cite{RoschRasch,StrohmaierGreif}. In our
simulations we do not see indications for a corresponding crossover
time scale $\sim 1/\gamma$ at which exponential relaxation might set
in. The power law decay in the Hubbard model (or, at least, the very
small relaxation rate) might be a consequence of the infinitely many
local conservation laws leading to integrability. It is important to
stress, however, that our numerical data for the intermediate time
dynamics cannot finally resolve the question whether or not
exponential relaxation does exist. In the next paragraph we will,
however, give further support that the fits with Eq.~\eref{fit_fct}
describe the relaxation at long times correctly by showing that the
the asymptotic value $d_U^D(\infty)$ obtained from the fits agrees
very well with a thermal expectation value.

\subsubsection{Long-time limit and thermalization}
According to the eigenstate thermalization hypothesis (ETH)
\cite{Deutsch, Srednicki}, each initial state---which can be represented as a
superposition of eigenstates of the Hamiltonian---already contains a
thermal state.  This thermal state is revealed during time evolution
due to dephasing effects between the different eigenstates. We say
that the system has thermalized if the long-time average
\begin{equation}
\label{long_t_ave}
\bar{o}=\lim_{\tau\to\infty} \frac{1}{\tau}\int_0^\tau dt \langle \Psi_I|o(t)|\Psi_I\rangle
\end{equation}
is equal to the thermal average $\langle o\rangle_{\rm \lambda_i}$ in
an appropriately chosen ensemble with the intensive variables
$\lambda_i$. Note that this definition only demands that
$\bar{o}=\langle o\rangle_{\rm \lambda_i}$, i.e., time dependent
fluctuations in $\langle \Psi_I|o(t)|\Psi_I\rangle$ can, in principle,
remain large even for $t\to\infty$. This will, in particular, be true
for free models where no relaxation mechanisms exist and the concept
of thermalization therefore has limited meaning. In interacting
models, on the other hand, we expect that $o(t\to\infty)\equiv
\bar{o}$, i.e., time-dependent fluctuations vanish in the long-time
limit. In this case, the system is expected to have truly thermalized
if $o(t\to\infty)=\langle o\rangle_{\rm \lambda_i}$.

The appropriate thermal ensemble is determined by the set of conserved
quantities $Q_i$ with $[H,Q_i]=0$. Obviously, $\langle
\Psi_I|Q_i(t)|\Psi_I\rangle=\mbox{const}$ which means that the
intensive variables (Lagrange multipliers) $\lambda_i$ have to be
determined such that
\begin{equation}
\label{Lagrange_multi}
\langle \Psi_I|Q_i|\Psi_I\rangle \equiv \langle Q_i\rangle_{\rm
  \lambda_i}.
\end{equation}
Every quantum system has two types of conserved quantities: local and
non-local. We call a conserved quantity local if it can be expressed
as a sum of local densities acting only on a finite number of lattice
sites.\footnote{Equivalently, a conserved quantity in a field theory
  is local if it can be written as an integral of a fully local
  operator density.} For a generic system, usually only very few local
conserved quantities such as the particle number operator and the
Hamiltonian itself exist. Any quantum system in the thermodynamic
limit has, on the other hand, infinitely many non-local conserved
quantities, as, for example, the projection operators
$|E_n\rangle\langle E_n|$ onto the eigenstates of the system. It is
not clear if these non-local conserved quantities play any role in
determining thermalization or transport
\cite{JungRoschDrude,SirkerPereira,SirkerPereira2}. In studies of
thermalization they are usually simply neglected.

The 1D Hubbard model is integrable by Bethe ansatz and has infinitely
many local conserved quantities which can be constructed explicitly
from a family of commuting transfer matrices
\cite{ShastryHubbard,ZotosPrelovsek}. In this case, one should, in
principle, consider a generalized Gibbs ensemble with a Lagrange
multiplier $\lambda_i$ for each local conservation law.  However, such
ensembles are impossible to handle in the thermodynamic limit except
for the simplest free particle models \cite{CassidyClark}. Here we
will instead consider the usual canonical ensemble. The effective
temperature $T_{\rm eff}$ then acts as a Lagrange multiplier
determined such that
\begin{equation}
\label{Teff}
\langle \Psi_D|H|\Psi_D\rangle/L = \frac{1}{LZ}\Tr\left\{ H\e^{-H/T_{\rm eff}}\right\}
\end{equation}
where $Z=\Tr\e^{-H/T_{\rm eff}}$ is the partition function and the
energy $\langle \Psi_D|H|\Psi_D\rangle/L=U/4$ is conserved during time
evolution. Since the spectrum of eigenenergies per site is bounded for
a lattice model, the use of negative temperatures is natural with
$1/T\to 0^{\pm}$ corresponding to the case of maximum entropy. In the
following we denote the thermal average in the canonical ensemble by
$\langle d\rangle_{U,T}$ and it is easy to see that $\langle
d\rangle_{U,T}=\langle d\rangle_{-U,-T}$ holds. The duality
transformation \eref{duality}, \eref{eq:dUD} furthermore implies
$\langle d\rangle_{U,T}+\langle d\rangle_{-U,T}=1/2$ for all $T$. In
particular, $\langle d\rangle_{U>0,T>0} <1/4$ and $\langle
d\rangle_{U<0,T>0} >1/4$ with both being equal to $1/4$ in the limit
$T\to\infty$. We calculate the thermal average $\langle
d\rangle_{U,T}$ using a transfer-matrix DMRG algorithm
\cite{BursillGehring, WangXiang, Nakamura, GlockeSirker} and find that
$T_{\rm eff}$ is negative (positive) for the repulsive (attractive)
model, respectively.
The dependence of the effective temperature $T_{\rm eff}$ on
interaction strength is shown in Fig.~\ref{Fig4b}(a) together with the
results for the extended Hubbard model which are discussed in the next
subsection.  The comparison between the double occupancy extrapolated
in time, $d^D_U(\infty)$, and the thermal double occupancy $\langle
d\rangle_{T_{\rm eff}}$ shown in Fig.~\ref{Fig4b}(b) yields excellent
agreement.
This seems to suggest, on the one hand, that the extrapolation using
Eq.~\eref{fit_fct} is appropriate leaving very little room for an
additional exponential decay which possibly could set in at a longer
time scale. On the other hand, it also seems to suggest that the other
local conservation laws of the Hubbard model have very little
influence on the relaxation of double occupancies. To qualitatively
understand the latter property we can think of writing the operator
$d$ as a sum of projections onto all the conserved quantities which
form a basis of the operator space
\cite{JungRoschDrude, SirkerPereira}. At least for large $U$ it is then
clear that the projection onto the Hamiltonian, which does contain the
operator $d$ itself, will give the dominant contribution to the
thermal expectation value.

\subsection{Results for the extended Hubbard model}
\label{Sec_ExtHubb}
\begin{figure}[!ht]
\begin{center}
\includegraphics*[width=0.6\columnwidth]{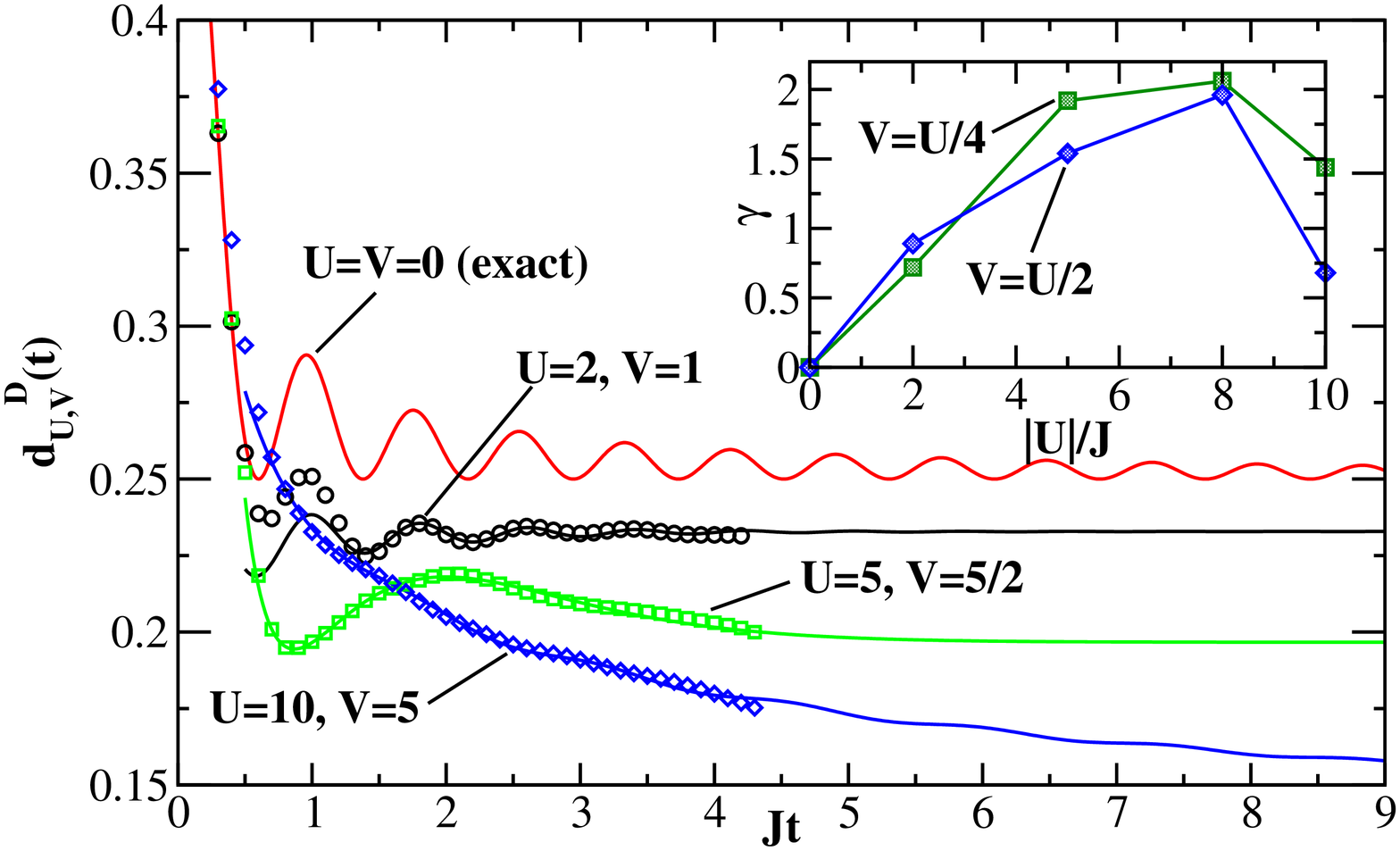}
\end{center}
\caption{Extended Hubbard model: $d^D_{U,V}(t)$ for $V=U/2$ with $\chi
  =10000$, $\delta t=0.1$ (symbols), and fits (lines). Inset:
  Relaxation rate $\gamma$ extracted from the fits.}
\label{Fig4}
\end{figure}

Finally, we consider the non-integrable {\it extended Hubbard model}
obtained by turning on the next-nearest neighbor repulsion $V$ in
Eq.~(\ref{Hubbard}).  The duality relations used for the Hubbard model
are then no longer valid because $V\sum_j (n_j-1)(n_{j+1}-1)\to
4V\sum_j S^z_j S^z_{j+1}$ under the duality tranformation,
Eq.~(\ref{duality}). However, we still have $d^D_{U,V}(t) =
d^D_{-U,-V}(t)$ for the time evolution and $\langle
d\rangle_{U,V,T}=\langle d\rangle_{-U,-V, -T}$ for the thermal
expectation value since these relations only rely on the lattice being
bipartite and time reversal symmetry.  We focus in the following on
$V\leq U/2$ corresponding to a spin-density wave state in the ground
state phase diagram \cite{Nakamura, GlockeSirker}. We note that for
$V>U/2$, $|\Psi_D\rangle$ is close to the charge-density wave (CDW)
ground state and long simulations in time are possible with
$d^D_{U,V}(t)$ staying close to $1/2$ and showing revival oscillations
(data not shown).  For the case $V=U/2$---which is approximately at
the phase transition line from the spin-density to a charge-density
wave state \cite{Nakamura, GlockeSirker}---we find a qualitatively different
behavior than in the Hubbard model (see
Fig.~\ref{Fig4}).
Here $d_{U,V}^D(t)$ {\it decreases} at long times with increasing
interaction strength. In general, $d_{U,V}^D(t)$ can increase or
decrease at long times with increasing interaction strength depending
on the ratio $V/U$.  Using the same fit function (\ref{fit_fct}) as
before we find that it is no longer possible to describe the
relaxation dynamics by a pure power law decay. Instead, we now
find a finite relaxation rate $\gamma$ as shown in the inset of
Fig.~\ref{Fig4}.

\subsubsection{Long-time limit and thermalization}
\begin{figure}[t!]
\begin{center}
\includegraphics*[width=0.7\columnwidth]{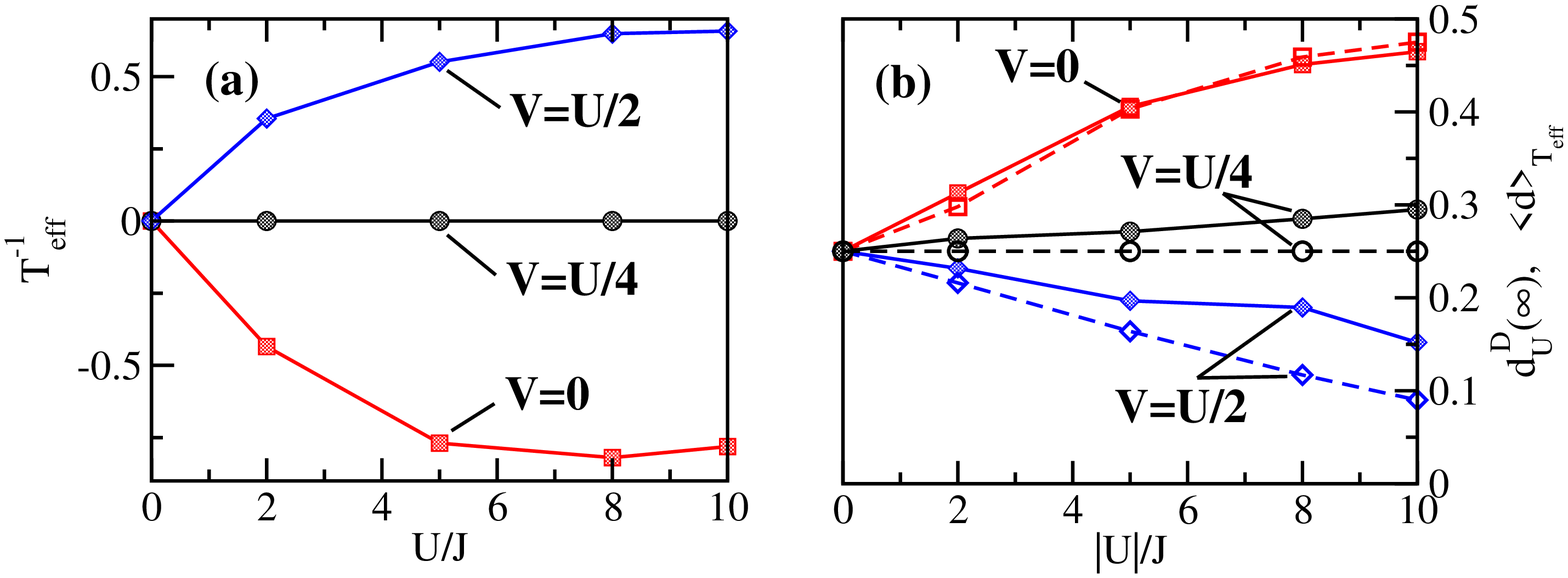}
\end{center}
\caption{(a) Inverse effective temperatures for different ratios of $V/U$ with
  $U>0$. $T_{\rm eff}$ changes sign for $U<0$. (b) $d^D_{U,V}(\infty)$
  (filled symbols connected by solid lines) compared to the thermal
  average $\langle d\rangle_{\rm T_{eff}}$ (open symbols connected by
  dashed lines).}
\label{Fig4b}
\end{figure}

Investigating again a possible thermalization we can shed some light
on the observed dependence of the extrapolated value
$d_{U,V}^D(\infty)$ on the ratio $V/U$. For $V\neq 0$ the model is no
longer integrable and---if thermalization does occur---the final state
should be fully described by the canonical ensemble.  The energy
during the time evolution is now fixed to
\begin{equation}
\label{innerE}
\langle \Psi_D|H|\Psi_D\rangle/L = \frac{U}{4}-V 
\end{equation}
and determines via the relation (\ref{Teff}) the effective temperature
shown in Fig.~\ref{Fig4b}(a). For $V=U/4$ it follows that $1/T_{\rm
  eff}=0$ and $\langle d\rangle_{U,V=U/4} =1/4$. In the repulsive
case, $U>0$, the energy (\ref{innerE}) is negative for $V>U/4$ leading
to $T_{\rm eff}>0$. For $V<U/4$, on the other hand, the energy is
positive and therefore $T_{\rm eff}<0$. For $U<0$ the signs of $T_{\rm
  eff}$ are reversed.
Using again a transfer matrix DMRG algorithm to calculate the thermal
expectation value $\langle d\rangle_{\rm T_{eff}}$ at temperatures
$T_{\rm eff}$ we can compare with the extrapolated value
$d^D_{U,V}(\infty)$ from the time evolution, see Fig.~\ref{Fig4b}(b).
Compared to the pure Hubbard model the agreement is not quite as good
and the deviations increase the closer the ratio of the interactions
is to the critical line $V=U/2$ and also the larger $U$ is. This does
suggest---assuming that the system will finally thermalize---that the
numerically obtained intermediate time dynamics is not sufficient to
fully extract the long time behavior. A possible explanation is that
two different relaxation processes exist in this case: a fast one at
short time scales leading to a pre-thermalized state and a slower one
setting in at $Jt\gg \exp(U/J)$ \cite{RoschRasch,StrohmaierGreif}.
This might also explain the non-monotonic behavior of
$\gamma(U,V/U=\mbox{const})$ obtained when fitting with a single
relaxation rate, see inset of Fig.~\ref{Fig4}.

\section{Application to a non-translationally invariant case}
\label{Sec_Non-trans}
\begin{figure}[t!]
\begin{center}
\includegraphics*[width=0.7\columnwidth]{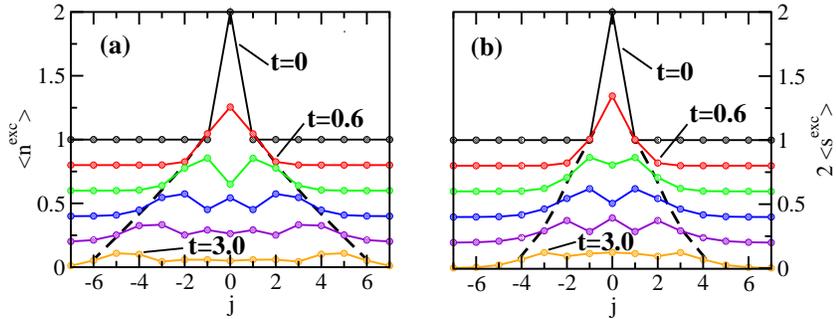}
\end{center}
\caption{Time evolution starting from the initial state
  $|\tilde\Psi_N\rangle$, Eq.~(\ref{Doublon_state}), for $U/J=2$ with
  $\chi=1024$ states kept. Shown are results for times
  $Jt=3,2.4,1.8,\cdots,0$ (from bottom to top) for (a) the excess
  charge $\langle n^{\rm exc}_j\rangle$, and (b) the excess spin
  $\langle s^{\rm exc}_j\rangle$.  Subsequent curves are shifted by
  $0.2$ for clarity of presentation. The dashed lines connect the
  points $x_{c(s)}$, see Fig.~\ref{Doublon_velocities}.}
\label{Doublon_dynamic}
\end{figure}

On of the main advantages of the LCRG algorithm is that it allows to
study the time evolution of one-dimensional quantum systems in the
thermodynamic limit even if the initial state and/or the Hamiltonian
are not translationally invariant. As an example, we consider the time
evolution in the $V=0$ Hubbard model \eref{Hubbard} of the
non-translationally invariant state
\begin{equation}
  |\tilde\Psi_N\rangle = c^\dagger_{0\uparrow}|\Psi_N\rangle
  =  c^\dagger_{0\uparrow}\prod_j c^\dagger_{2j+1\uparrow}c^\dagger_{2j\downarrow}|0\rangle
\label{Doublon_state}
\end{equation}
obtained by adding an additional electron at site $j=0$ to the N\'eel
state, Eq.~(\ref{states}). In Fig.~\ref{Doublon_dynamic}, LCRG results
for the dynamics of the excess charge and the excess spin
defined by
\begin{equation}
 \langle n^{\rm exc}_j\rangle = \langle n_j\rangle - \langle n_j^{\rm
   bg}\rangle \, ; \qquad
 \langle s^{\rm exc}_j\rangle = \langle s_j\rangle - \langle s_j^{\rm bg}\rangle
\label{excess}
\end{equation}
are shown for $U/J=2$. Here $\langle n_j^{\rm bg}\rangle\equiv 1$
($\langle s_j^{\rm bg}\rangle$) are the background charge (spin)
densities, respectively, obtained from the time evolution of
$|\Psi_N\rangle$, i.e., from a system without the additional electron.
Videos of the time evolution for several other interaction strengths
$U/J$ are presented in the supplementary material. We observe that
$\langle n^{\rm exc}_j\rangle$ and $\langle s^{\rm exc}_j\rangle$
spread out into the lattice with different velocities clearly
revealing the light-cone structure. For the non-translationally
invariant problem considered here we have not implemented the
conservation laws yet and the results shown in
Fig.~\ref{Doublon_dynamic} have been obtained by keeping a relatively
moderate number of states, $\chi=1024$. Although the simulation time
is therefore smaller than in the translationally invariant cases
discussed in the previous chapters, we want to stress that the
evolution of the N\'eel background is simulated in the thermodynamic
limit and thus very different from that in a small system tractable,
for example, by exact diagonalization.

Different charge and spin velocities for the one-dimensional Hubbard
model starting from an initial non-equilibrium state have already been
observed in Ref.~\cite{KollathSchollwoeck}. In this case, an initial
state was considered where the ground state was perturbed by a small
local charge and spin imbalance thus allowing to extract the
velocities of the elementary spin and charge excitations which can
also be obtained by Bethe ansatz. Our initial state, on the other
hand, is a highly excited state and the charge and spin velocities are
not related to those of the elementary excitations.

To extract the charge velocity $v_c$ and the spin velocity $v_s$ for
our initial state we take the point $x_{c(s)}(t)$ where the tail
has reached half of the height of the first peak of the charge (spin)
distribution as reference point.
\begin{figure}[t!]
\begin{center}
\includegraphics*[width=0.7\columnwidth]{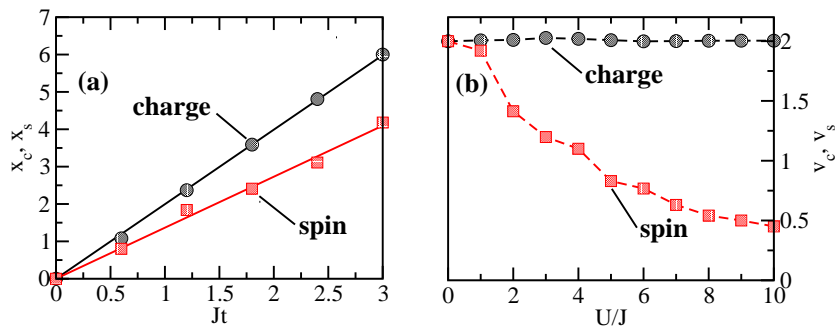}
\end{center}
\caption{(a) Time evolution of the reference points $x_{c(s)}$ as
  defined in the text for the excess charge (circles) and for the
  excess spin (squares) with $U/J =2$. The lines are linear fits. (b)
  Charge and spin velocities as a function of $U/J$.}
\label{Doublon_velocities}
\end{figure}
Fig.~\ref{Doublon_velocities}(a) shows that $x_{c(s)}(t)$ depends
linearly on time. The dependence of the velocities on the interaction
strength $U/J$ is shown in Fig.~\ref{Doublon_velocities}(b). We find a
charge velocity $v_c\approx 2J$ independent of $U$. For $U=0$ the
charge and spin distributions are identical for all times and
$v_s=v_c$, but for increasing $U$ the spin velocity $v_s$ decreases.
These results can be qualitatively understood as follows: the excess
charge sees a uniformly charged background and therefore moves
unimpeded with the Fermi velocity $v_F$ of the non-interacting system,
$v_c\approx v_F=2J$. The excess magnetization, on the other hand,
moves with a velocity proportional to the effective spin superexchange
$\sim J^2/U$ which therefore decreases with increasing $U$. For large
$U$ we find, furthermore, that the spin dynamics becomes more
complicated. While most of the excess magnetization remains inert in
this limit, an additional small staggered part appears which spreads
out with a velocity close to the charge velocity (data not shown).

\section{Conclusions}
\label{Conc}
The development of new spectroscopic techniques to study ultracold
quantum gases on optical lattices with very good spatial and time
resolution has put the topic of non-equilibrium dynamics in quantum
systems firmly back onto the agenda. Particularly interesting from a
fundamental persepective is the dynamics in one dimension where many
of the standard lattice models such as the Heisenberg, the $t-J$, and
the Hubbard model are integrable, i.e., these systems have an infinite
number of local conservation laws. Since a conserved quantity stays
invariant under time evolution, the presence of many conserved
quantities is expected to severely restrict the dynamics of the
quantum system as a whole and might even prevent the system from
reaching thermal equilibrium. To investigate such questions
numerically, different algorithms have been developed:  The
time-dependent density-matrix renormalization group (t-DMRG) and the
time-evolving block decimation (TEBD), in particular, allow one to
access the intermediate time dynamics by representing the time evolved
state as a matrix product state. Among the aims when developing
numerical algorithms to study the time evolution are (a) the
thermodynamic limit, (b) the flexibility to simulate different
systems, (c) a high computational efficiency, and (d) to simulate the
system for as long as possible.

Here we have presented a new algorithm which does make progress
concerning many of the points mentioned above. The light cone
renormalization group (LCRG) algorithm is highly efficient by
simulating at each time step only that part of the lattice (a light
cone) which is affected by the time evolution. The results obtained
are directly for the thermodynamic limit. Contrary to the infinite
size TEBD, the LCRG algorithm does not rely on translational
invariance. It is therefore extremely flexible and can, in particular,
also deal with non-translationally invariant problems. It cannot,
however, simulate the quantum system for times significantly longer
than other algorithms based on matrix product states. We have shown
that the lattice path integral representation of unitary time
evolution---forming the basis for the LCRG algorithm---provides a
simple picture for the linear entanglement growth with time which
restricts the simulation time of such algorithms. An executable sample
code for the anisotropic Heisenberg model is provided in the
supplementary material.

We used the LCRG algorithm to study quench dynamics in the integrable
Hubbard model starting from a state with every second site doubly
occupied and found indications for a pure power-law relaxation.  For
the extended non-integrable Hubbard model, on the contrary, the
relaxation appears to be exponential. In both cases we found that the
time evolved state in the long-time limit seems to be close to a
thermal state supporting the eigenstate thermalization hypothesis.

Finally, we demonstrated that the LCRG algorithm can also be used to
study the time evolution of non-translationally invariant initial
states. For the N\'eel state with one site occupied by a doublon we
showed that the excess charge and excess spin spread with finite, but
different, velocities. We extracted the dependence of the velocities
on the strength of the Hubbard interaction $U$ which can be used for
comparison with future experiments. Videos of the time evolution for
different parameter sets can be found in the supplementary material.

As an outlook we want to emphasize that future applications of the
LCRG algorithm to other non-translationally invariant setups such as
impurity problems, disorder, and to systems with trapping potentials
are feasible.

\ack The authors thank P.~Barmettler and A. Rosch for valuable
discussions.  J.S. acknowledges support by the graduate school of
excellence MAINZ/MATCOR and by the DFG via the SFB/Transregio 49.

\section*{References}

\providecommand{\newblock}{}

\end{document}